\documentclass[preprint,twocolumn,10pt,superscriptaddress, aps,pre]{revtex4-2}
\usepackage{amsmath,amssymb,amsfonts,csquotes}
\usepackage{graphicx}
\usepackage{xcolor}
\usepackage{float}

\usepackage{hyperref}
\hypersetup{
colorlinks=true,
urlcolor= blue,
citecolor=blue,
linkcolor= blue,}

\usepackage{bm}

\begin{document}
\title{Modeling contagious disease spreading  }
\author{Dipak Patra}

\email[]{dipak@rri.res.in}
\affiliation{Soft Condensed Matter Group, Raman Research Institute, C. V. Raman Avenue, Sadashivanagar, Bangalore-560080, India
 }
\date{\today}
\begin{abstract}
An understanding of the disease spreading phenomenon based on a mathematical model is extremely needed for the implication of the correct policy measures to contain the disease propagation.
 Here, we report a new model namely the Ising-SIR model describing contagious disease spreading phenomena including both airborne and direct contact disease transformations. In the airborne case, a susceptible agent can catch the disease either from the environment or its infected neighbors whereas in the second case, the agent can be infected only through close contact with its infected neighbors. We have performed Monte Carlo simulations on a square lattice using periodic boundary conditions to investigate the dynamics of disease spread. The simulations demonstrate that the mechanism of disease spreading plays a significant role in the growth dynamics and leads to different growth exponent. In the direct contact disease spreading mechanism, the growth exponent is nearly equal to two for some model parameters which agrees with earlier empirical observations. In addition, the model predicts various types of spatiotemporal patterns that can be observed in nature.
 
\end{abstract}

\maketitle

\section{Introduction}
The spreading of contagious diseases, a ubiquitous phenomenon in nature, is irrespectively found across demographic areas including villages, towns, countries, and even the globe where the unconstrained transmission can impact society in a multitude of ways resulting in a threat to public health and an economic slowdown. Therefore, the understanding of the spreading mechanism is much more needed to contain the disease transmission and uphold economic growth. However, the inherent complexity of the phenomenon often makes understanding a challenging task and profoundly establishes this phenomenon as an interdisciplinary subject of research \citep{Ross_1916, Grassly_2008}. To reduce the complexity of the system, some simple dynamical rules are often assumed in the mathematical modeling. These simple assumptions are primarily tested against the observational data to validate the model and its prediction. Sometimes a simple model can predict the robust outcomes giving rise to a better insight into the phenomena. Nevertheless, the prediction helps the policymakers to take effective measures such as lockdown, quarantine, social distancing, and vaccination, to prevent disease propagation \citep{Glasser_2004}. Therefore, the mathematical modeling of disease spreading has been studied over the years, decades, and even centuries \citep{Ross_1916, Kermack_1927, Coburn_2009, Chowell_2016, Mello_2021, Arenas_2020}. In particular, the number of studies adopting mathematical models has surged over the last four years because of the COVID-19 pandemic \citep{Arenas_2020}. Despite a large number of studies, finding out the appropriate mathematical model is an extremely difficult task and it is an active field of research.

The susceptible -- Infectious -- Recovered (SIR) model, a most useful deterministic model for epidemiological studies, was first proposed by Kermack and McKendrick in 1927 \citep{Kermack_1927}. In this model, the population is categorized into three compartments such as susceptible (S), infectious (I), and recover (R) states. These three states correspond to the basic dynamical variables in any infectious disease spreading. At any instant, the values of these states are determined by solving the ordinary differential equations constructed with a few assumptions. Over the past four years, most of the studies related to the COVID-19 disease have been carried out using this model because of its simplicity and computational cost efficiency. Depending on the macroscopic parameters the model predicts the different types of temporal growth profiles. 
 Similar types of SIR models with an additional number of compartments have been employed to describe the disease spreading.
 However, the implementation of these models is limited in real systems due to their apparent simplicity as these models neglect the stochastic nature of the systems, inhomogeneity over space, and finally microscopic details of the agents. 
Some of these constraints can be removed by incorporating the noise and diffusive terms in the dynamical equations of the SIR model where the dynamics of disease spreading are governed by the partial differential equations \citep{te_Vrugt_2024}. 
However, these types of models are also limited to the incorporation of the microscopic properties of agents and their interactions. Hence, the role of microscopic properties on the dynamics of disease spreading has been explored using agent-based models.

Active Brownian particle model, network model, and cellular automata are commonly known as agent-based models. In these models, the agents are mostly classified into three classes S, I, and R. For the Brownian model, the movement of the agents is governed by the stochastic partial differential equations which are analogous to the Langevin equations describing active matter systems \citep{Norambuena_2020, Kailasham_2024}. In this model, the disease spreading occurs through the movement of agents and the interactions with their infected neighbors. In the network model agents are represented by the nodes of a graph. The infected nodes spread the disease to their next susceptible nodes which are directly connected by the edges in the network \citep{Browne_2021}. In the cellular automata model, square or hexagonal lattices are often chosen as a system of interest wherein each lattice point can be described as either an empty or a filled cell  \citep{White_2007}. Each filled cell in the lattice represents the agents of the population. The susceptible cell can get the disease through contact with infected neighbor-filled cells. In all these models, the S-I transformation is governed by some probabilistic rule whereas the I-R transformation is controlled by either the deterministic or probabilistic rules. Hence, these dynamic rules dictate the dynamics of disease spreading.
 In most of the earlier studies, the cellular automata model 
 has been employed because of its easier numerical implementation and computational cost efficiency compared to the other two models.

Another physics-based model with Ising spins, describing a population divided into two classes, has also been studied to understand the temperature role in the diffusion of the COVID-19 virus focusing on the seasonal effect on the pandemic \citep{Serra_2021}. In this model, the Ising spin represents either the S or I state of an agent and the interaction between two agents is strongly regulated by the Ising-like Hamiltonian in contrast to the cellular automata model. Nevertheless, a quantum version of the Ising model has been employed to study the COVID-19 pandemic \citep{Padhi_2020}. However, both these physics-based studies are not only limited to the population with two compartments but also lack an understanding of the spatiotemporal growth dynamics of the disease spreading.  

Here, we have studied the contagious disease spreading dynamics by developing an Ising-SIR model. The Ising spin accounts for the three states S, I, and R corresponding to the disease spreading and the spins interact with each other with the proposed Hamiltonian. We have performed the Monte Carlo simulations for both airborne and direct contact disease transmission mechanisms. Our model demonstrates the temporal as well as spatial profile of growth dynamics. Different types of spatial patterns are observed and their formations are strongly dictated by the mechanism of disease transmission. For the airborne infectious disease, the patterns consist of infectious patches in the system whereas the circular patterns are obtained for the contact infectious disease. Nevertheless, the temporal growth profiles for these two cases are completely different. For some model parameters, the growth exponent for the contact infectious disease is nearly equal to two which agrees well with the earlier empirical observations. 

\section{ Ising-SIR Model}
Susceptible, infectious, and recovered classes are considered the basic three compartments in the population for any contagious disease spreading. Thus each agent remains in any of these three states at any moment. These states can be ascribed by the Ising spin 1 that is denoted by $s$ and the spin states +1, 0, and -1  represent the susceptible, infectious, and recovered state of an agent, respectively. It should be noted that the agent can change its state by interacting with its neighbor and environment. For theoretical and computational studies of a physical system, the Hamiltonian, generally defined by incorporating these interactions, dictates the dynamics of the system. Therefore, at a minimal level, we construct the Hamiltonian
\begin{equation}
H=-Jk_{B} \Big[\sum_{<(i,j)>} (s_i s_j)^2 -D\sum_i s^2_i \Big]
\end{equation}
for the system describing the spread of a contagious disease where $< >$ includes the nearest neighbor interaction and the sums run over the full system. For a physical system  $J$ is a constant of the dimension of temperature and $k_B$ is the Boltzmann constant. Here, $J$  is assumed to be positive in the system. 
The first term attains the minimum value of $-1$ for S-S, S-R, and R-R pairs of agents. It favors the mixing of susceptible and recovered agents accounting for the coexistence and cooperative behavior among these two classes in society. Other pairs such as S-I, I-I, and R-I lead to the maximum value. These pairs are therefore unlikely to occur in the ideal or disease-free society. The second term favors the infectious state of an agent for $D>0$. The parameter $D$ either relates to the contamination strength of a disease or the immunity of an agent in a reciprocal way. A similar type of model with three spin states known as the Blume-Emery-Griffiths (BEG) model has been used to describe a physical system focusing on phase transition and phase separation in He$^3$-He$^4$ mixtures where $D$ is called crystal field representing the strength of anisotropy in the spin state \citep{Blume_1971}.
 
Monte Carlo simulations over a square lattice of size $200\times 200$ with periodic boundary conditions have been employed for understanding the disease spreading dynamics. Some constraints are imposed on the system depending on the mechanism of disease propagation. The Metropolis algorithm is used to update the spin state in simulations with these constraints.
The spin updating schemes for two methods of disease spreading are described in the following. 

 For the airborne transmission process, we assume that the disease spread occurs through the air as well as through the nearest-neighbor infectious contact.
 Initially, all the agents are supposed to be in the susceptible ($s=1$) state.
 The spin updating scheme is described by the following steps:
  (i) An agent is randomly picked over the square lattice. The spin state of the agent is denoted by $s_r$. 
 (ii) A possible new spin state of the agent is  $s_n$. The value of $s_n$ strongly depends on the value of $s_r$ resulting in three conditions being discussed below. 
If the agent is in the susceptible state ($s_r = 1$) then the new state would be in the infectious state ($s_n=0$).
  This condition ensures the transition from susceptible to infectious states ( $S \rightarrow I $). In the second condition, the new state would be in the recovered state ($s_n=1$) if the agent is in the infectious state ($s_r = 0$). This condition implies a recovery procedure ( $I \rightarrow R $).
Lastly, the new spin is in the recovered state ($ s_n=-1$) for the recovered agent ($ s_r=-1$). Therefore, an individual in a recovery state remains in the same state during the simulation time. 
All these conditions describe the unidirectional $S \rightarrow I \rightarrow R$ transitions. 
  (iii)  The energy corresponding to the selected agent is calculated by using the proposed Hamiltonian and is denoted by $\epsilon_r$. Again, the calculation of energy associated with the new spin $\epsilon_n$ is performed by selecting $s_n$ as the agent spin. The energy difference  $\Delta \epsilon =  \epsilon_n - \epsilon_r$ is determined.
 (iv) The agent selects $s_n$ as the final state if the energy difference becomes less than zero ($\Delta \epsilon <0$). 
 For other cases, a random number ($u$) is drawn from a uniform random distribution (0,1).  The agent changes its state from $s_r$ to $s_n$ for the condition $e^{-\frac{\Delta \epsilon}{ k_B T}} > u$.  In the physics-based system, $T$ represents the absolute temperature. 
 (v) All the above steps (i)-(iv) are repeated for $200\times 200$ iterations which is considered to be one MC step. 
 All the simulations are run for a desired number of MC steps to probe the dynamics of disease spreading. It should be noted that the time unit for our simulation is one MC step.
 \begin{widetext}
\begin{minipage}{\linewidth}
\begin{figure}[H]
   \includegraphics[width=17cm]{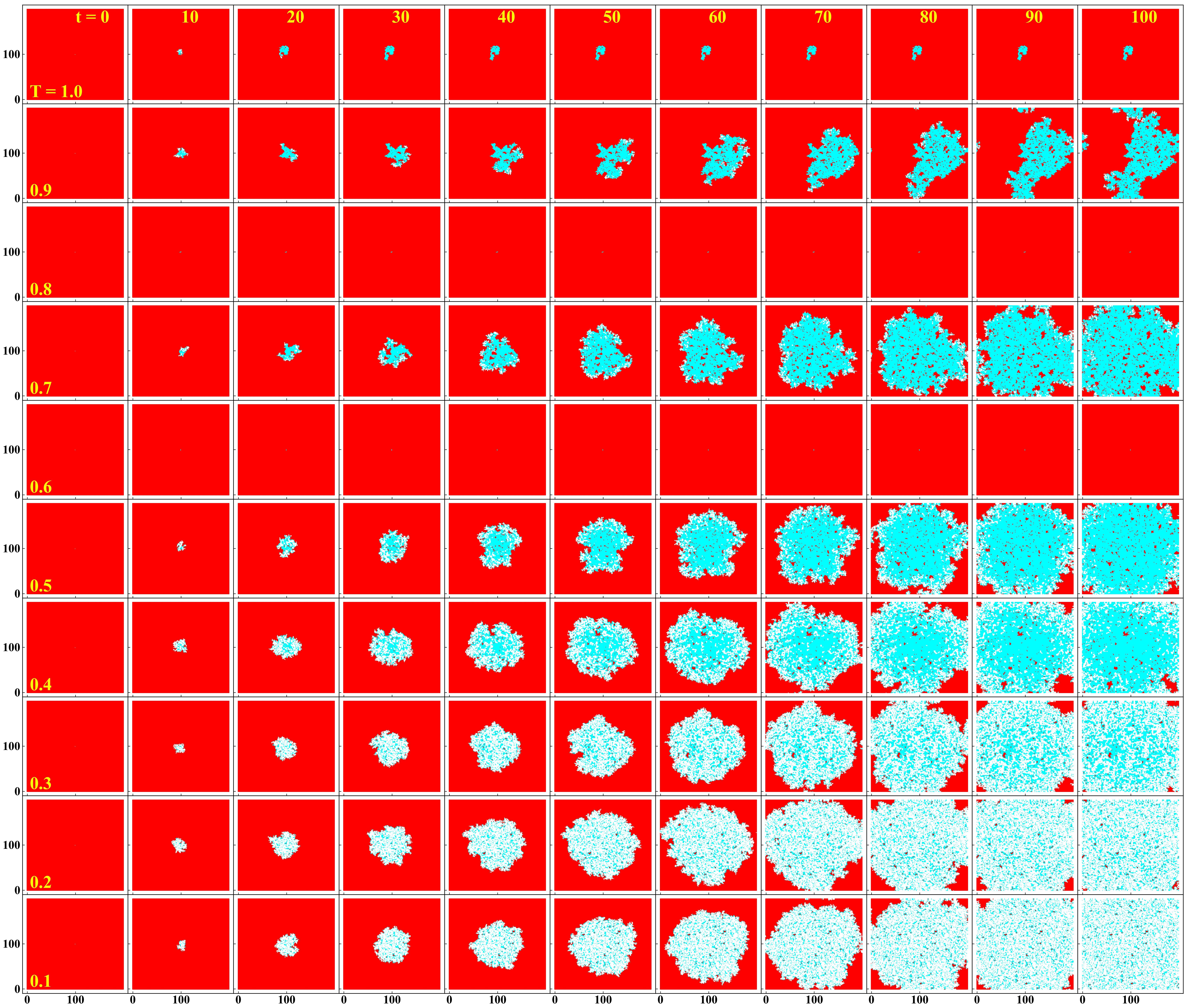}
   \centering
    \caption{Snapshots of the time evolution of the spin configuration with direct contact transmission for $D=3$ and different values of $T$ where red, white, and cyan colors representing the susceptible, infectious, and recovered agents respectively.} 
    \label{fig:DirectConfig}
\end{figure}
\end{minipage}
\end{widetext}

 For the direct contact disease transmission process, the disease spread occurs only through direct contact with the nearest neighbor infectious agent. At the starting point of the simulation, an infectious agent is chosen at the center of the lattice to spread the disease in the system. The rest of the spins are expected to be in the susceptible state. The update scheme for the spin variable more or less follows the above-discussed algorithm for the airborne disease transmission with a slight modification of step (ii):
 If $s_r = 1$ then the total number of infectious neighbors surrounding the randomly selected agent is counted. Then, the new spin state of the selected agent $s_n =0 $ is chosen for any nonzero number of infectious neighbors. Otherwise, the new spin is expected to be in the susceptible state $s_n =1 $.

Disease spreading dynamics is assumed to be an unidirectional process ($S \rightarrow I \rightarrow R$). Therefore, this unidirectional process leads to the breaking of detailed balance conditions and makes the system out of equilibrium. Hence, these simulations are different from the usual equilibrium MC simulations of statistical physics problems. Henceforth, a dimensionless model parameter $T/J$ is redefined as $T$ without any loss of generality. 
 This dimensionless parameter can be associated with many societal things such as anxiety, distress, disorder, and agitation among the agents. Hence, it can be called a turmoil parameter.
 \begin{widetext}
\begin{minipage}{\linewidth}
\begin{figure}[H]
   \includegraphics[width=17cm]{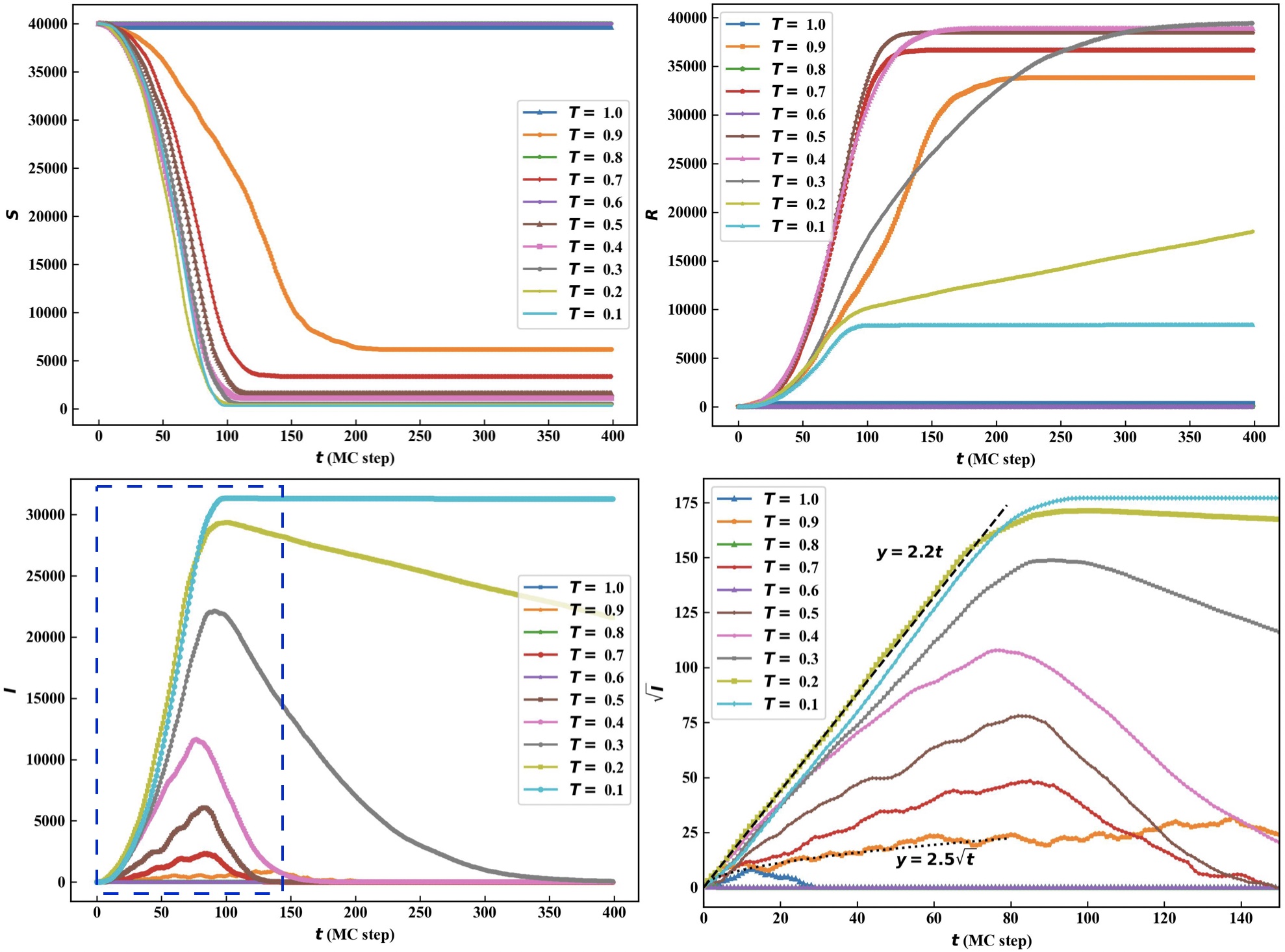}
   \centering
    \caption{Epidemic curves for the individual runs presented in Fig.\ref{fig:DirectConfig} where (a), (b) and (c) displaying S, I and R respectively. (d) Presenting the temporal variation of the square root of I corresponding to the marked region in pan~(c). The dashed and dotted lines are qualitatively drawn to confirm power-law growth dynamics in I.} 
    \label{fig:DirectSIR}
\end{figure}
\end{minipage}
\end{widetext}
\section{Results and discussions}
\begin{figure}[ht]
  \centering 
   \includegraphics[clip=true,width=\columnwidth]{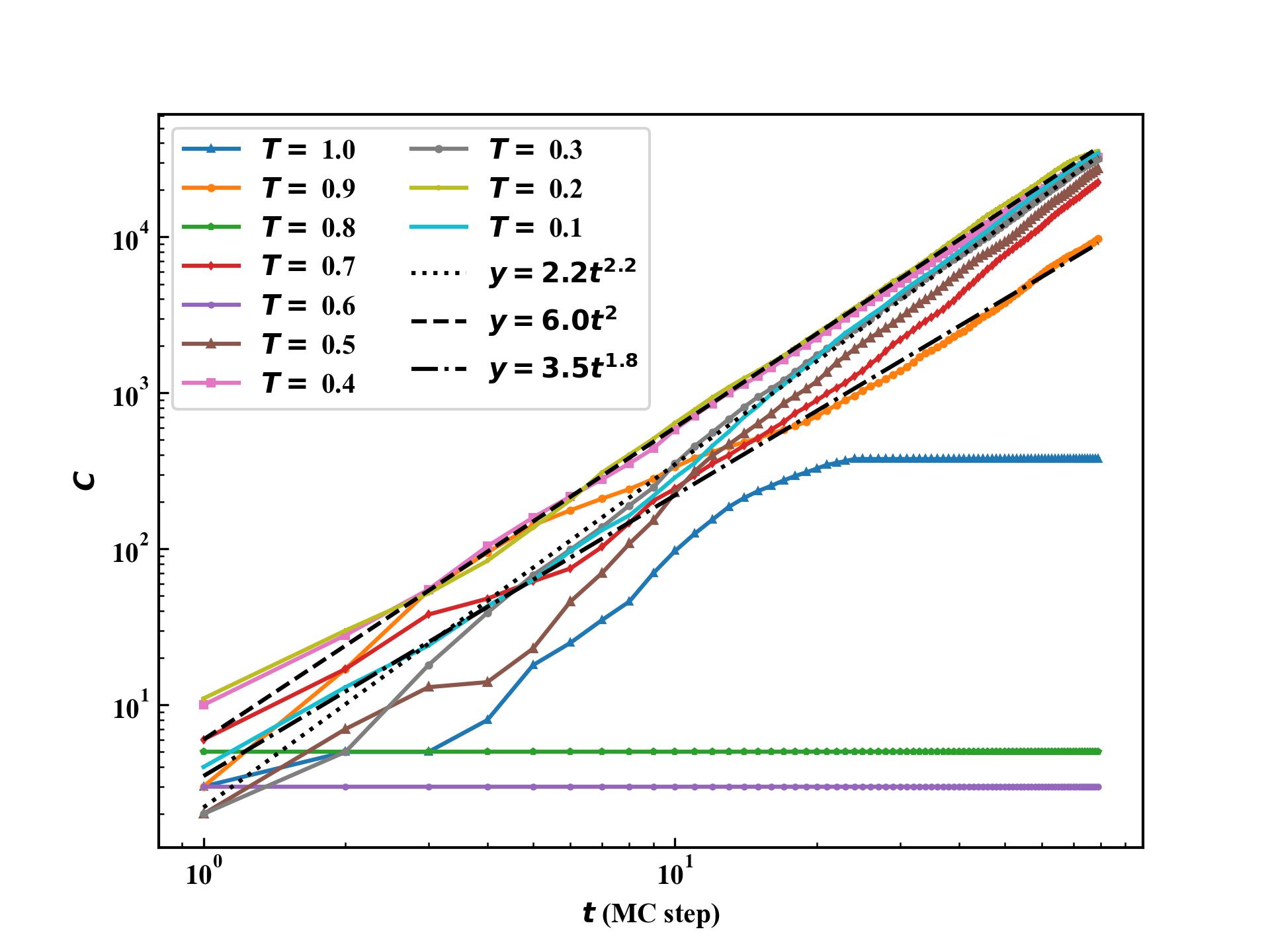}
   \caption{ The variation of the confirmed cases $C$ as a function of time
corresponding to the individual runs presented in Fig.\ref{fig:DirectConfig}. The dashed, dotted, and dashed-dotted lines are qualitatively drawn to confirm power-law growth dynamics.}
    \label{fig:DirectCumulative}
\end{figure} 
We have performed Monte Carlo simulations to investigate the effects of transmission mechanisms on disease spreading. Figure \ref{fig:DirectConfig} displays the typical spatiotemporal patterns corresponding to the direct contact spreading mechanism for various values of turmoil parameter $T$. Each of these patterns is nucleated from a single infectious agent located at the center of the square lattice. For higher values of $T$, patterns grow in time with an anisotropic growth front where most of the infectious and recovered agents are found to be located near the boundary and inside of the patterns respectively. Whereas for lower values of $T$ the patterns exhibit almost a circular growth front where recovered and infectious agents are not segregated. It has been found that after a large number of time steps some susceptible agents remain inside the pattern without catching the disease and they are surrounded by the recovered agents. This type of behavior has been empirically observed in disease spreading phenomena in a society which is often described by hard immunity. Temporal variations of the number of agents in three states corresponding to these patterns are presented in figure \ref{fig:DirectSIR}.  The number of susceptible agents mostly decreases with time and can attain saturation values at large times as shown in figure \ref{fig:DirectSIR} (a). Whereas the number of infectious agents initially increases with time then attaining a maximum value it starts to decrease and reaches zero at large time (see figure \ref{fig:DirectSIR} (b) ). As can be seen in figure~\ref{fig:DirectSIR} (c), the number of recovered agents increases from zero with time and can attain a saturation value by ensuring the conservation of the total number of agents. These typical nature of growth dynamics in $S, I$, and $R$ have been found in the conventional SIR model. For higher values of $T$, the transition rate from $I\rightarrow R$ is higher compared to that of for lower values of $T$. It is therefore found that the maximum number of infectious agents increases with decreasing values of $T$. However, the time corresponding to the peak position does not significantly vary with $T$. It has been found that the number of infectious agents exhibits power law growth dynamics (i.e. $I\sim t^\mu$ where $\mu$ is the growth exponent) in the earlier stage of spreading dynamics. To find out the growth exponent $\mu$ for $I$, we present the variation of the square root of the number of infectious agents as a function of time in figure \ref{fig:DirectSIR} (d).
For lower values of $T$, $I$ varies quadratically with time i.e. $I\sim t^2$ with the exponent $\mu = 2$ as the disease spreads with a more or less circular growth front and the infectious agents are almost everywhere in the patterns (see figure~\ref{fig:DirectConfig} for $T=0.1$ and $0.2$). The number of infectious agents is therefore more or less proportional to the area of the pattern. It should be noted that the radius/size of the pattern grows linearly with time giving rise to the growth exponent equal to $2$. For higher values of $T$ the infectious agents remain on the periphery of the patterns (see figure~\ref{fig:DirectConfig} for $T=0.7$ and $0.9$) and therefore it grows linearly i.e. $I\sim t$ with the exponent $\mu =1$. For intermediate values of $T$, the infectious agents are found to be located near the circumference with a finite width as can be seen in figure~\ref{fig:DirectConfig} for $T=0.4$. 
Hence, the growth exponent for these cases is expected to lie in a range $1\leq \mu \leq 2$. The knowledge of $\mu$ for $I$ would be beneficial to contain the disease propagation as it strongly depends on the position of infectious agents. However, most of the previous studies have focused on the growth dynamics for the cumulative number of infectious agents ($C=I+R$) at the earlier stage of disease propagation and found the power-law behavior with a growth exponent $\nu$ as $C \sim t^\nu$. The growth exponent $\nu>=1$ has been observed in varieties of diseases such as Ebola, influenza, smallpox, plague measles, and HIV/AIDS \citep{Colgate_1989, Viboud_2016}. Interestingly similar types of growth dynamics have been reported in the Covid-19 pandemic where the exponent mostly equals $2$ \citep{ Maier_2020, Brandenburg_2020}. In this case, the observation of slower growth dynamics has been explained as the effects of measures against the disease such as lockdown/restriction in the movement of the citizen. Notably, diseases with direct contact transmission such as HIV/AIDS also exhibit $\nu=2$. Figure~\ref{fig:DirectCumulative} presents the variation of $C$ as a function of time with a logarithmic scale for both axes. The growth exponent $\nu$ is found to be near equal to $2$. 
It should be noted that the preferential growth model can explain the observation of $\nu=2$. However, the exponent less than $2$ observed in our model arises due to the anisotropic pattern with a fractal-like boundary. Whereas the exponent greater than $2$ observed in our model arises due to the preferential attachments with additional growth dynamics inside of the pattern.
It should be noted that for any values of $T$ the system can exhibit only susceptible and recovered agents after a few time steps resulting in the end of the disease spreading that arises due to the stochastic nature of MC simulations under the constraint of the close contact transmission (see figure~\ref{fig:DirectConfig} and \ref{fig:DirectSIR} for $T=1.0,0.8$ and $0.6$). To obtain the average statistics, we have performed $50$ independent MC runs for each value of parameter $T$. The average statistics are presented in the appendix section which also provides the same findings discussed above (see figure~\ref{fig:SuppDirectSIR} and figure~\ref{fig:SuppDirectCumulative}).

Now we present the typical spatiotemporal patterns corresponding to the airborne spreading mechanism for various values of turmoil parameter $T$ in figure~\ref{fig:AirConfig}. For higher values of $T$, the system exhibits a random mixture of three states $S, I$, and $R$. Whereas for lower values of $T$, the disease spreading occurs with domain-like structures. These domains continuously grow similar to the direct contact transmission process as long as they do not collide with each other. In these cases, the total number of domains varies with time. The variation of the total number of susceptible, infectious, and recovered agents as a function of time corresponding to these spatiotemporal patterns are presented in figure~\ref{fig:AirSIR} (a), (b), and (c) respectively. Similar types of temporal profiles for these three states are generally observed in the conventional SIR model.
The number of susceptible agents starts to decrease rapidly in contrast to the direct contact transmission process. At large time the system is not expected to contain any susceptible agent. Similarly, $I$ increases rapidly, and after reaching a maximum value it decreases with time. The time taken by the system to reach a disease-free state decreases with increasing values of $T$. The number of recovered agents increases with time and reaches the saturation value (i.e. total number of agents). The time taken to reach the saturation level increases with decreasing values of $T$. 
 \begin{widetext}
\begin{minipage}{\linewidth}
\begin{figure}[H]
   \includegraphics[width=17cm]{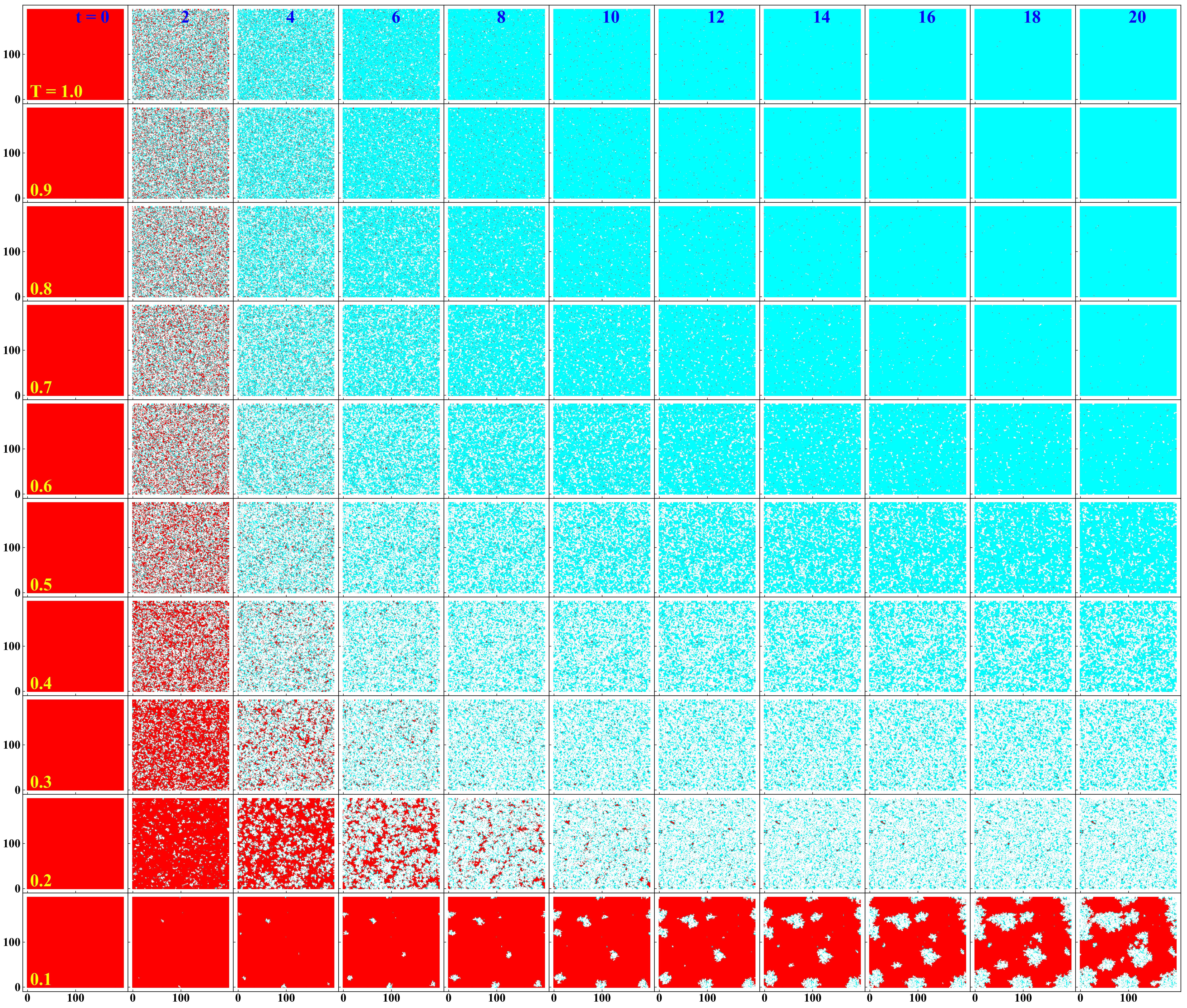}
   \centering
    \caption{Snapshots of the time evolution of the spin configuration corresponding to the airborne disease transmission for $D=3$ and different values of $T$ where red, white, and cyan colors representing the susceptible, infectious, and recovered agents respectively.} 
    \label{fig:AirConfig}
\end{figure}
\end{minipage}
\end{widetext}

\begin{figure}[t]
  \centering 
   \includegraphics[clip=true,width=\columnwidth]{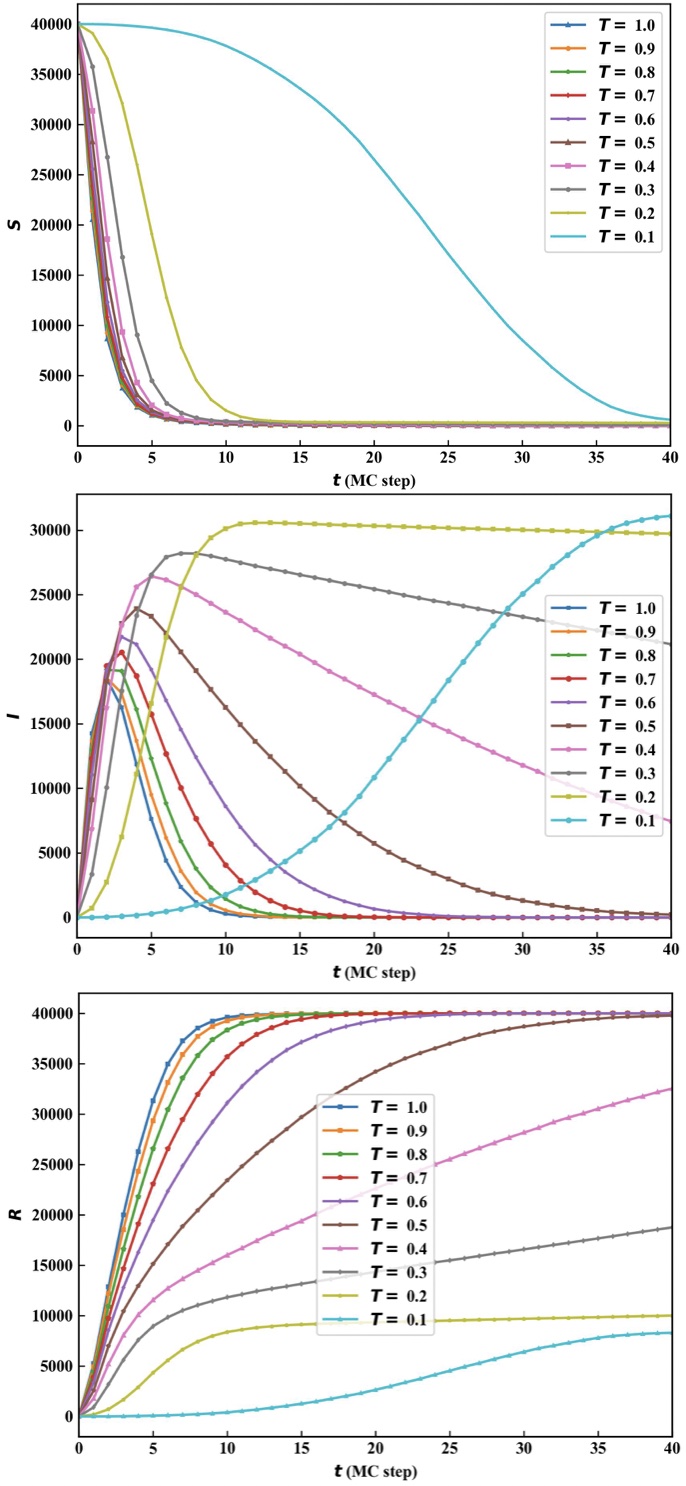}
   \caption{Epidemic curves for the individual runs presented in Fig.\ref{fig:AirConfig} where (a), (b) and (c) displaying the profiles of S, I, and R respectively. }
    \label{fig:AirSIR}
\end{figure}
\begin{figure}[h]
  \centering 
   \includegraphics[clip=true,width=\columnwidth]{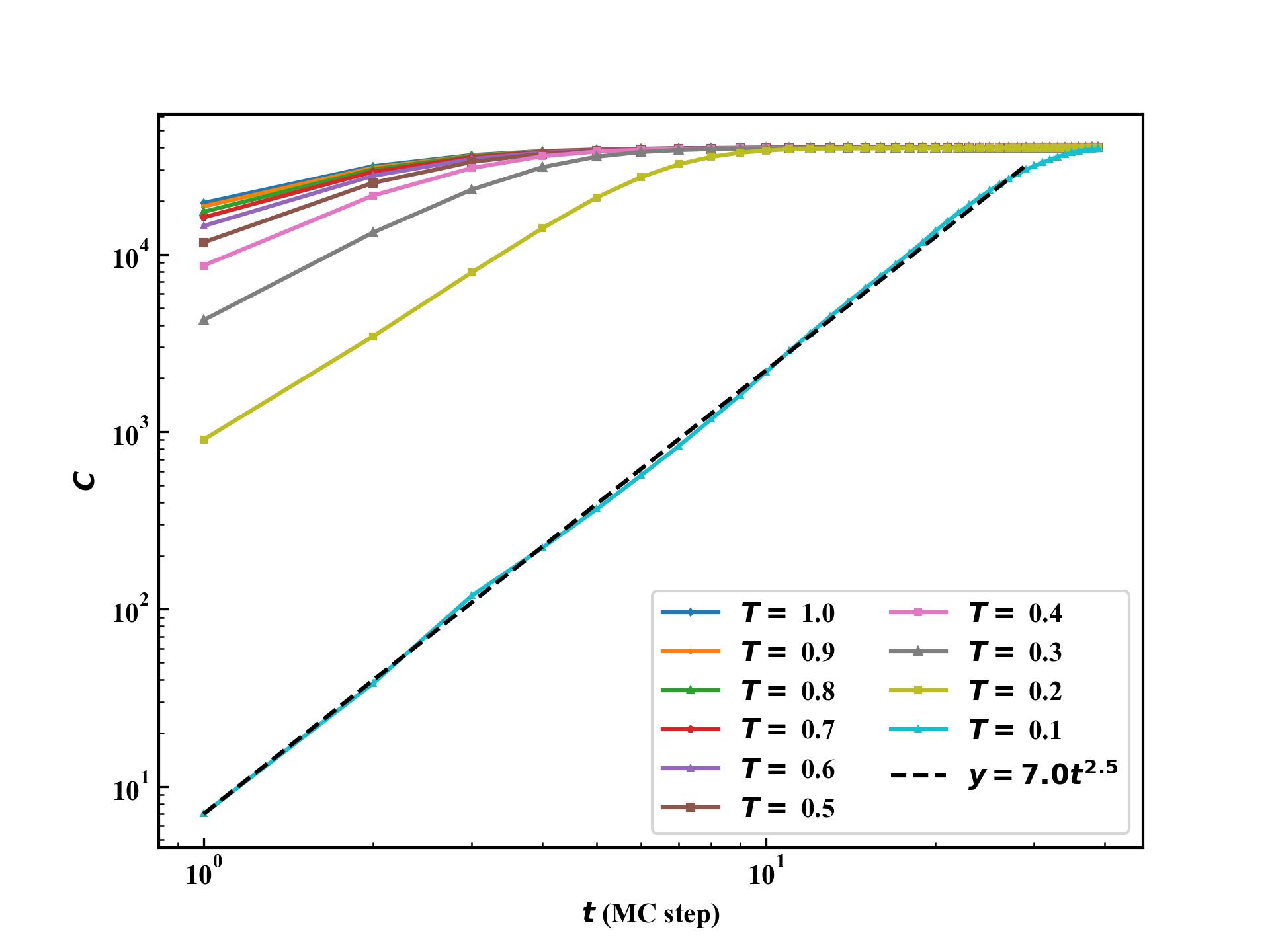}
   \caption{The variation of the confirmed cases $C$ as a function of time
corresponding to the individual runs presented in Fig.\ref{fig:AirConfig}. The dashed line is qualitatively drawn to confirm power-law growth dynamics. }
    \label{fig:AirCumulative}
\end{figure}

  
Figure~\ref{fig:AirCumulative} presents the variation of the cumulative number of infectious agents $C$ as a function time for different values of model parameter $T$. The $C$ exhibits the complementary behavior compared to $S$ as the total number of agents is a conserved quantity. For higher values of $T$, the cumulative number $C$ reaches the saturation value within a short span of time giving rise to the exponential growth. Whereas for lower values of $T$, the cumulative number $C$ increases nonlinearly with time exhibiting power-law growth dynamics as $C\sim t^\nu$. In these cases, the domain-like structures mostly spread the disease following the direct contact mechanism giving rise to the power-law growth dynamics. For $T=0.1$, the growth exponent $\nu$ is greater than $2$. The spontaneous creation of the domains leads to a higher growth exponent. It can be thought of in such a way that an infectious agent from a domain moves to another part of the system and spreads the disease by creating a new domain \citep{Triambak_2020}. To obtain the average statistics, we have performed $50$ independent MC runs for each value of parameter $T$. The average temporal profiles of these three states are presented in the appendix section and these profiles are almost identical to the respective profiles discussed above (see figure~\ref{fig:SuppAirSIR} and figure~\ref{fig:SuppAirCumulative}).

It should be noted that the movement of agents plays an important role in any infectious disease spreading. Following earlier literature, the impact of movement can be studied in this model by allowing some empty cells in the square lattice and an agent can move to the empty cell depending on some dynamical rules. Nevertheless, multi-compartmental models have also been studied in the literature for describing the various observations in the Covid-19 pandemic. To conduct similar studies our model needs to be extended for a higher number of states by including Pott-like interaction among the agents. The spin system with a larger number of states has been considered in the Pott model for describing the critical phenomenon in statistical physics. It should be noted that the birth and death of an agent can impact on the disease spreading that have not been considered in the present study. The effect of changes in the total number of agents on the spatio-temporal dynamics of disease spreading is aimed for future studies. 
We have not thoroughly compared our findings with the survey data related to disease propagation. However, our model qualitatively supports the findings reported in earlier literature. We, therefore, intend to study a detailed comparison with empirical observations in future studies. Nevertheless, further investigation is needed to find out other scopes and limitations of our model. 

\section{Summary }
We have developed an Ising-SIR model for describing the contagious disease-spreading phenomenon. The Monte Carlo simulations of this model account for both airborne and contact transmission of the disease.
The spatiotemporal patterns and growth dynamics are mostly different for these two transmission modes. The circular growth front and growth exponent near equal to two are observed only for the contact mode of transmission. The simulation results qualitatively agree with the earlier empirical observations.

\onecolumngrid
\appendix
\section{Average statistics}

\begin{figure*}
   \includegraphics[width=17cm]{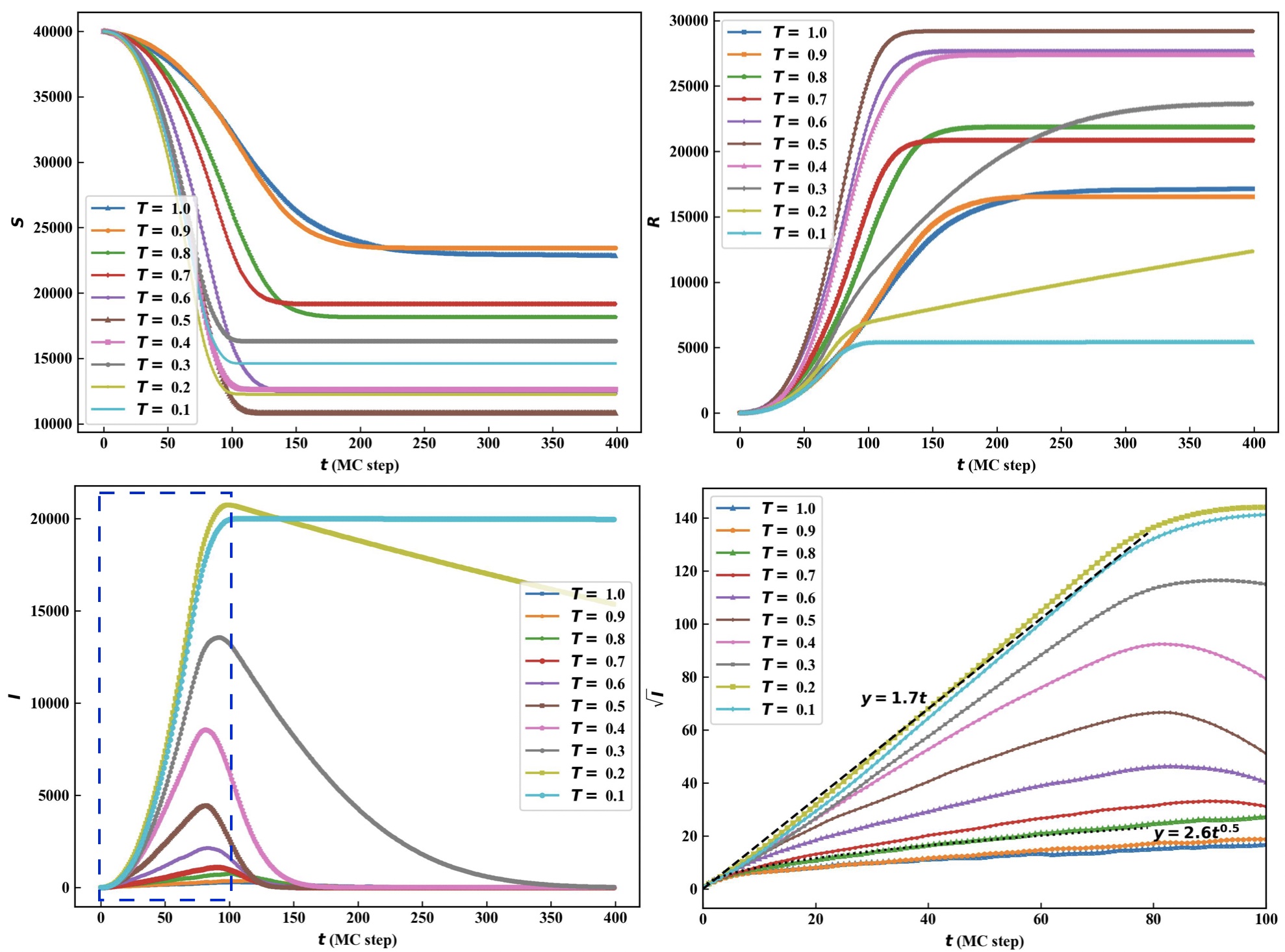}
   \centering
    \caption{Epidemic curves corresponding to the direct contact mechanism obtained by averaging over 50 independent MC runs where (a), (b), and (c) displaying the profiles of S, I, and R respectively. (d) presenting the temporal variation of the square root of I corresponding to the marked region in pan~(c). The dashed and dotted lines are qualitatively drawn to confirm power-law growth dynamics in I.} 
    \label{fig:SuppDirectSIR}
\end{figure*}

\begin{figure*}
  \centering 
   \includegraphics[clip=true,width=\columnwidth]{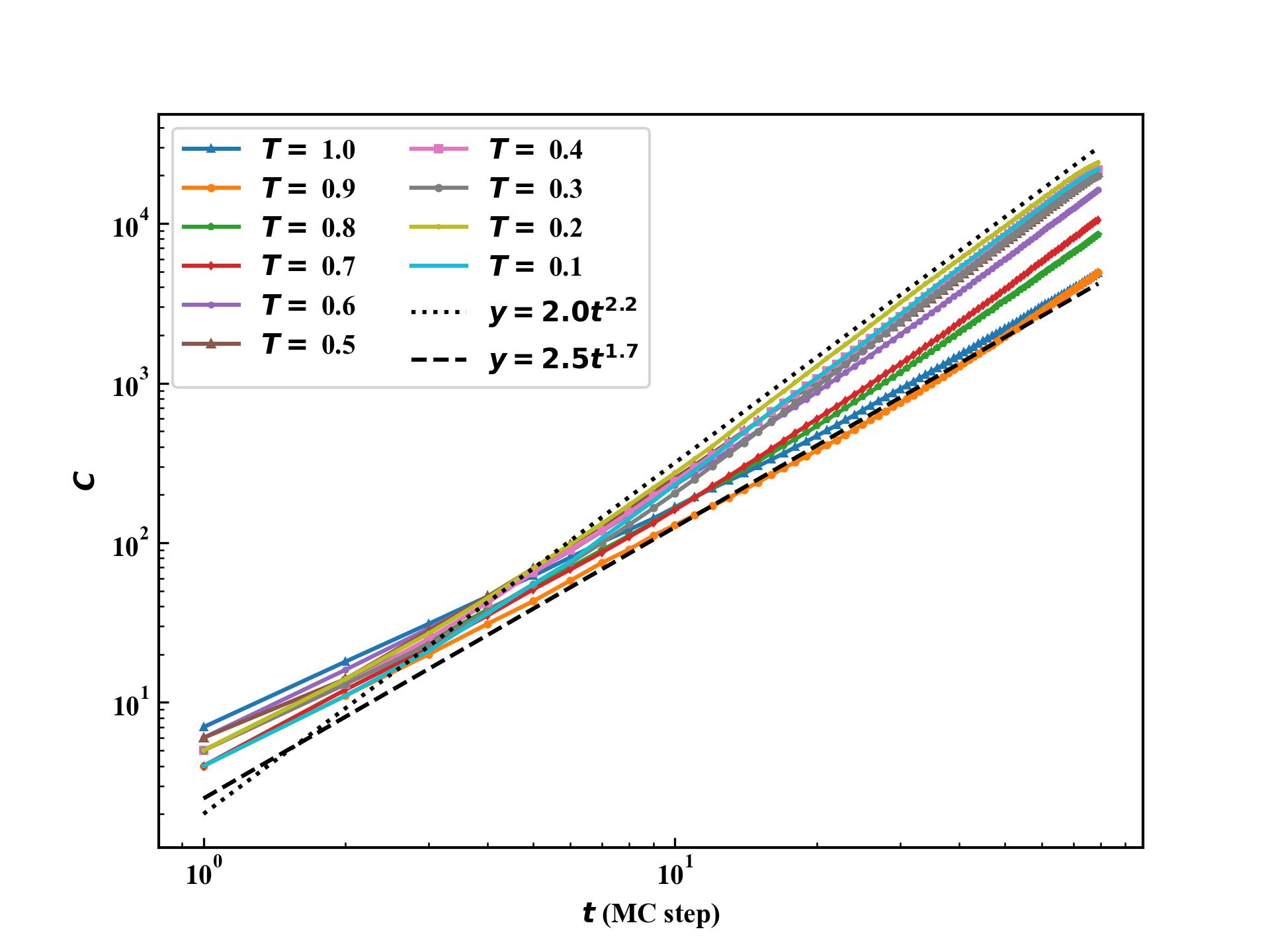}
   \caption{The variation of the confirmed cases $C$ as a function of time obtained by averaging over 50 independent MC runs corresponding to the direct contact mechanism. The dashed and dotted lines are qualitatively drawn to confirm power-law growth dynamics.}
    \label{fig:SuppDirectCumulative}
\end{figure*}

\begin{figure*}
   \includegraphics[width=17cm]{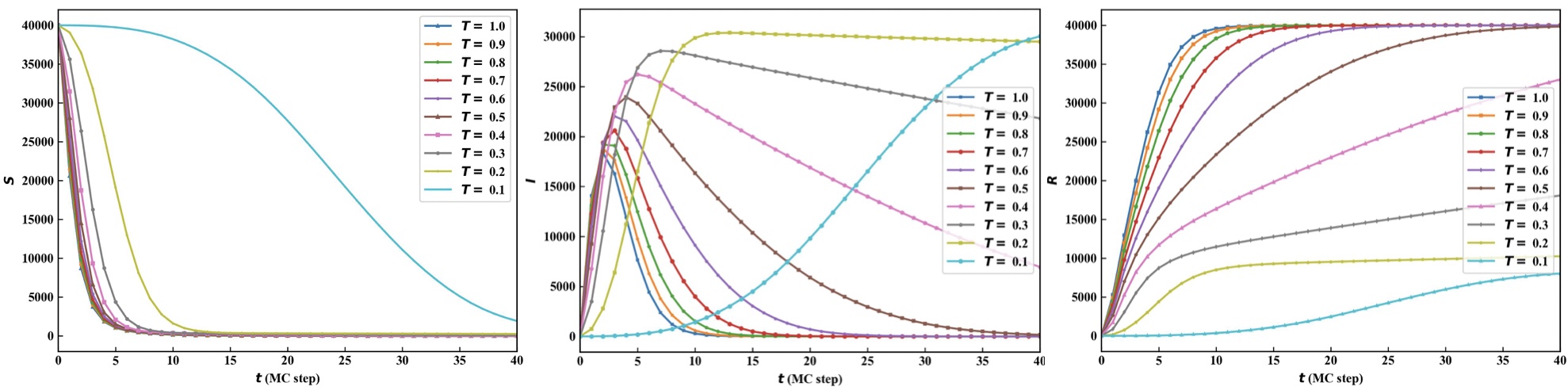}
   \centering
    \caption{Epidemic curves corresponding to the airborne mechanism obtained by averaging over 50 independent MC runs where (a), (b), and (c) displaying the profiles of S, I, and R respectively. } 
    \label{fig:SuppAirSIR}
\end{figure*}

\begin{figure*}
  \centering 
   \includegraphics[clip=true,width=\columnwidth]{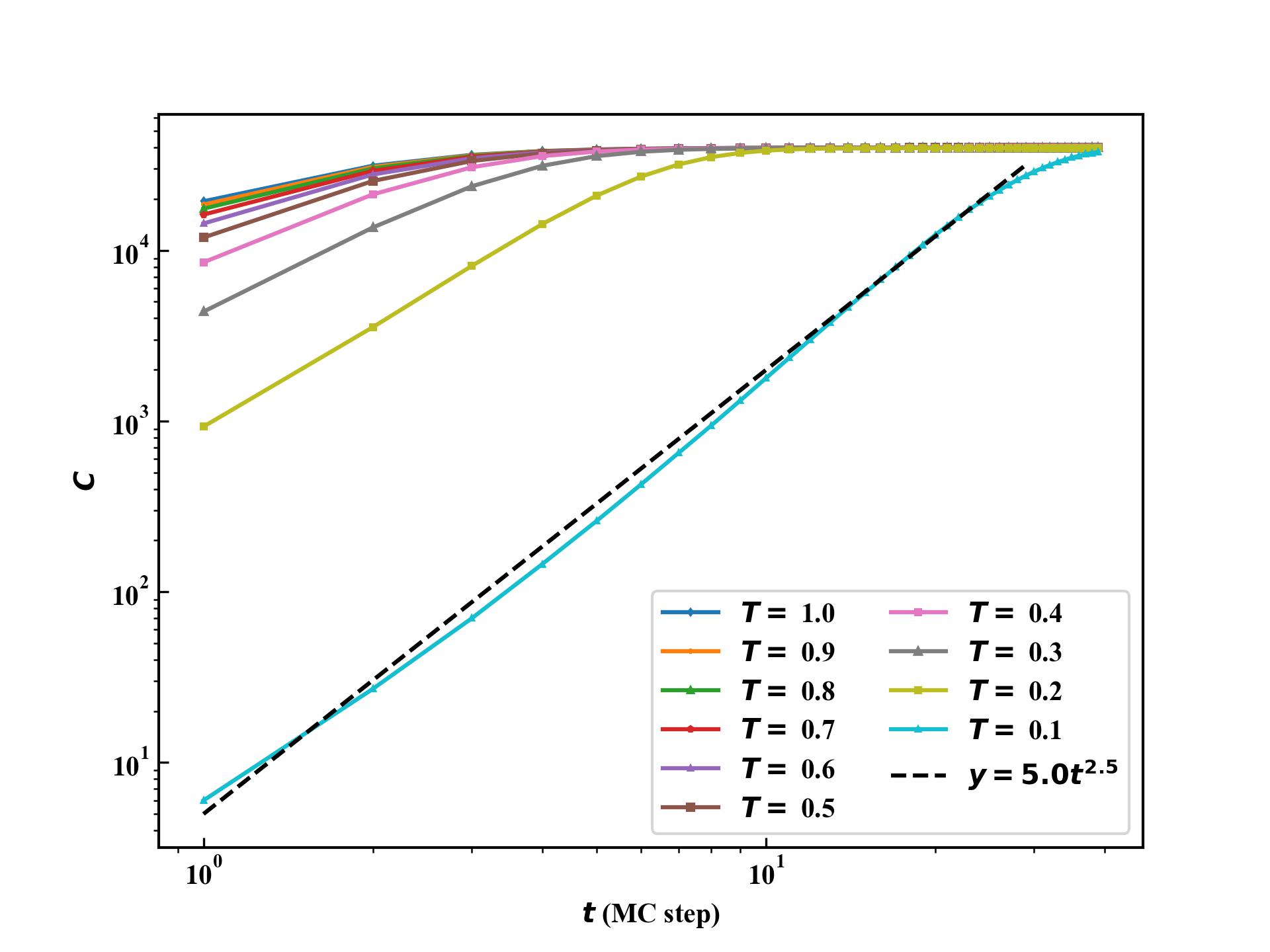}
   \caption{ The variation of the confirmed cases $C$ as a function of time obtained by averaging over 50 independent MC runs corresponding to the airborne mechanism. The dashed line is qualitatively drawn to confirm power-law growth dynamics.}
    \label{fig:SuppAirCumulative}
\end{figure*}

\end{document}